\documentclass[11pt,a4paper]{article}

\usepackage{graphicx,amsmath,natbib,bm,url,chngcntr}

\usepackage[hidelinks]{hyperref}
\hypersetup{
colorlinks=false,
linktocpage=false,
breaklinks=true}

\usepackage[utf8]{inputenc} 
\usepackage{graphicx}

\hypersetup{
colorlinks=darkred,
hypertexnames=false,
colorlinks,		
linkcolor={darkred},
citecolor={darkred},
urlcolor={darkred},
pdfstartview={FitV},
unicode,
breaklinks=darkred
} 

\usepackage[australian]{babel} 
\usepackage[a4paper,text={14.5cm, 25.2cm},centering]{geometry} 
\usepackage{mathpazo} 

\usepackage{longtable, booktabs, tabularx} 
\usepackage[flushleft]{threeparttable}  
\usepackage{subcaption}
 
\usepackage[onehalfspacing]{setspace} 

\usepackage[marginal]{footmisc} 
\usepackage{soul}

\usepackage{pstricks,pst-plot}
\usepackage{theorem}
\definecolor{darkred}{HTML}{690000}
\definecolor{nicegreen}{HTML}{556B2F}

\theoremheaderfont{\bfseries \upshape}

\usepackage{xr}
\usepackage{pgfplots}
\pgfplotsset{compat=1.13}
\usepackage
{todonotes}

\usepackage{tikz}
\usetikzlibrary{backgrounds, intersections,patterns}

\newcommand{\theoremref}[1]{\hyperref[#1]{Theorem~\ref*{#1}}}
\newcommand{\obsref}[1]{\hyperref[#1]{Observation~\ref*{#1}}}
\newcommand{\lemmaref}[1]{\hyperref[#1]{Lemma~\ref*{#1}}}
\newcommand{\defref}[1]{\hyperref[#1]{Definition~\ref*{#1}}}
\newcommand{\corref}[1]{\hyperref[#1]{Corollary~\ref*{#1}}}
\newcommand{\propref}[1]{\hyperref[#1]{Proposition~\ref*{#1}}}
\newcommand{\assref}[1]{\hyperref[#1]{Assumption~\ref*{#1}}}
\newcommand{\remref}[1]{\hyperref[#1]{Remark~\ref*{#1}}}
\newcommand{\exref}[1]{\hyperref[#1]{Example~\ref*{#1}}}
\newcommand{\exsref}[1]{\hyperref[#1]{Examples~\ref*{#1}}}
\newcommand{\appref}[1]{\hyperref[#1]{Online Appendix~\ref*{#1}}}
\newcommand{\secref}[1]{\hyperref[#1]{Section~\ref*{#1}}}
\newcommand{\subsecref}[1]{\hyperref[#1]{Subsection~\ref*{#1}}}
\newcommand{\figref}[1]{\hyperref[#1]{Figure~\ref*{#1}}}
\newcommand{\tableref}[1]{\hyperref[#1]{Table~\ref*{#1}}}
\newcommand{\procref}[1]{\hyperref[#1]{Procedure~\ref*{#1}}}

\newtheorem{Def}{\color{black}Definition}
\newtheorem{Prop}{\color{black}Proposition}

\newtheorem{Lem}{\color{black}Lemma}
\newtheorem{Cor}{\color{black}Corollary}
\newtheorem{Ass}{\color{black}Assumption}

\makeatletter  
\def\@endtheorem{\hfill $\diamond$\endtrivlist\@endpefalse } 
\makeatother

\theoremstyle{definition}
\newtheorem{Rem}{\color{black}Remark}

\DeclareMathOperator{\supp}{supp}

\setcitestyle{authoryear}

\begin{document}

\newcommand{\type}{\ensuremath{t}} 
\newcommand{\smin}{\ensuremath{\underline{\type}}} 
\newcommand{\smax}{\ensuremath{\overline{\type}}} 
\newcommand{\cscale}{\ensuremath{\frac{1}{\gamma}}} 
\newcommand{\typeh}{\ensuremath{\hat{\type}}}
\newcommand{\typeb}{\ensuremath{\type^{\prime}}} 
\newcommand{\Types}{\ensuremath{T}} 
\newcommand{\ac}{\ensuremath{a}} 
\newcommand{\acb}{\ensuremath{\ac^{\prime}}} 
\newcommand{\Ac}{\ensuremath{A}} 
\newcommand{\sig}{\ensuremath{\sigma}} 
\newcommand{\sigd}{\ensuremath{\tilde{\sig}}} 
\newcommand{\fals}{\ensuremath{\rho}} 
\newcommand{\Fals}{\ensuremath{R}} 
\newcommand{\csa}{\ensuremath{\cscale c(\ac,\type)}} 
\newcommand{\csab}{\ensuremath{\cscale c(\ac,\typeb)}} 
\newcommand{\cxa}{\ensuremath{\cscale  c(\ac,z)}} 
\newcommand{\cxang}{\ensuremath{c(\ac,z)}} 

\newcommand{\Cd}{\ensuremath{D}} 
\newcommand{\alocm}{\ensuremath{q_{\mu}}} 
\newcommand{\aloc}{\ensuremath{q}} 
\newcommand{\alocs}{\ensuremath{\bar{\aloc}}} 
\newcommand{\Sn}{\ensuremath{S}} 
\newcommand{\sn}{\ensuremath{s}} 
\newcommand{\snb}{\ensuremath{\sn^{\prime}}} 
\newcommand{\nmin}{\ensuremath{\underline{\sn}}} 
\newcommand{\nmax}{\ensuremath{\overline{\sn}}} 
\newcommand{\cscalem}{\ensuremath{\kappa}} 
\newcommand{\cscalemb}{\ensuremath{\cscalem^{\prime}}}
\newcommand{\typem}{\ensuremath{\sn,\cscalem}} 
\newcommand{\gmin}{\ensuremath{\underline{\cscalem}}} 
\newcommand{\gmax}{\ensuremath{\overline{\cscalem}}} 
\newcommand{\Gam}{\ensuremath{\mathcal{K}}}
\newcommand{\typemb}{\ensuremath{\sn^{\prime},\cscalem^{\prime}}} 
\newcommand{\csam}{\ensuremath{\cscalem c(\ac,\sn)}} 
\newcommand{\csamb}{\ensuremath{\cscalem c(\ac,\snb)}} 

\title{\textcolor{darkred}{Score-based mechanisms\\
\textsc{Preliminary and incomplete}}\thanks{Acknowledgements to be added. Eduardo Perez-Richet acknowledges funding by the European Research Council (ERC) consolidator grant 101001694. Vasiliki Skreta acknowledges funding by the NSF grant 2242521``Fraud-proof Mechanism Design''. 
}}

\author{
 Eduardo \textsc{Perez-Richet}\footnote{Sciences Po, CEPR -- e-mail: {\tt eduardo.perez@sciencespo.fr} }
\and 
Vasiliki \textsc{Skreta }\footnote{UT Austin, UCL, CEPR -- e-mail: {\tt vskreta@gmail.com}}}

\date{\today}

\maketitle \setcounter{page}{1}


\begin{abstract}
\setstretch{1}
 We propose a mechanism design framework that incorporates both soft information, which can be freely manipulated, and semi-hard information, which entails a cost for falsification. The framework captures various contexts such as school choice, public housing, organ transplant and manipulations of classification algorithms. We first provide a canonical class of mechanisms for these settings. The key idea is to treat the submission of hard information as an observable and payoff-relevant action and the contractible part of the mechanism as a mapping from submitted scores to a distribution over decisions (a score-based decision rule). Each type report triggers a distribution over score submission requests and a distribution over decision rules. We provide conditions under which score-based mechanisms are without loss of generality. In other words, situations under which the agent does not make any type reports and decides without a mediator what score to submit in a score-based decision rule.  We proceed to characterize optimal approval mechanisms in the presence of manipulable hard information. In several leading settings optimal mechanisms are score-based (and thus do not rely on soft information) and involve costly screening. The solution methodology we employ is suitable both for concave cost functions and quadratic costs and is applicable to a wide range of contexts in economics and in computer science.
\\
\medskip 

\noindent\textsc{Keywords:} Mechanism Design, Fabrication, Ordeals, Manipulation, Cheating, Costly screening.

\smallskip \noindent\textsc{JEL classification:} C72; D82.
\end{abstract}

\clearpage
\tableofcontents
 
\onehalfspacing
\newpage
\onehalfspacing

\section{Introduction}

\paragraph{Scoring Mechanisms}
Financing decisions, admissions to selective higher education institutions and allocations of public housing units and of human organs are often performed via \textit{score-based mechanisms.} These mechanisms rely on scores or priorities  that measure and aggregate agents' characteristics into a one-dimensional score or metric. For instance, in consumer finance, credit scores assess an individual's creditworthiness and determine loan terms. In the financial sector, stress test scores evaluate the health of financial institutions. Education institutions use standardized test scores to evaluate student performance, while public school seats are allocated via  priorities based on factors such as proximity to the school or a sibling in the school. In the medical field, human organ allocations are prioritized based on health status and treatments. 

Oftentimes scores or priorities do not merely reflect agents' natural or true characteristics due to manipulations. As \cite{frankel2019muddled} point out, this leads to a distinction between an agent's natural score--obtained without interfering with the measuring technology--and the measured score, which may result from gaming, manipulation, or falsification. In school choice settings, there is ample evidence that some families submit fake addresses to achieve entry in desirable schools.\footnote{For example, suggestive evidence indicates that parents fake addresses to gain admission to desirable public schools in Denmark \citep{bjerrenielsen2023playing}.} Doctors put their patients on escalated treatments in order to increase their priority on organ waiting lists \citep{lee2018manipulation,mcmichael2022stealing}.  There is ample evidence of gaming of classification algorithms as discussed, for example, in \cite{braverman2020role} and in the survey of \cite{Tang2023}. Manipulations can be costly and the cost depends on hard information  (agents' natural scores) and soft information (agents' abilities or tastes). Manipulations can impact the fairness and efficiency of these systems (see \citealp{hu2019disparate}).


We embark in \secref{sec-model} by proposing a unified framework that accommodates both soft and hard information. Traditionally in mechanism design, types represent soft information, allowing agents to lie freely. However, when introducing evidence, types are hard information and the standard assumption is that an agent either has a piece of evidence so the cost of lying is  zero or does not have it in which case the cost of misrepresentation is infinite. By contrast, we allow for richer evidence structures captured by general falsification costs. We proceed to provide in \secref{sec-can-mech} a canonical class of mechanisms for these settings. The key idea is to treat the submission of hard information as an observable and payoff-relevant action and the contractible part of the mechanism as a mapping from submitted scores to a distribution over decisions (a score-based decision rule). Each type report triggers a distribution over score submission requests and a distribution over score-based decision rules. This allows us to map this setting with costly misreporting to one captured by the generalized principal-agent setting in \cite{myerson1982optimal} and obtain that truthful and obedient direct recommendation mechanisms (DRMs) are without loss of generality.

Beyond the revelation principle, in our setting, we obtain two further simplifications that hold in general. First, DRMs can be decomposed into two mappings:  each type report is mapped to a single \textit{score-based decision rule} which, in turn, maps publicly submitted scores to a distribution over decisions and a \textit{score recommendation rule} (a.k.a. falsification strategy) that maps a type report to  random score recommendations. In other words, the mechanism does not need to randomize over score recommendation rules; randomization may be needed for score recommendations only. Second, because the agent submits scores publicly, obedience constraints are implied by voluntary ex-post participation constraints. 

While the revelation principle in our setting identifies the language of inputs and outputs of the mechanism (type reports and score submission recommendations respectively) it does not specify that agents should be recommended to submit their natural scores. It is possible that optimal mechanisms request agents to submit falsified scores\footnote{For an example, optimal tests in the presence of costly falsification leverage, what \cite{perez2022test} coin \textit{productive falsification}, to improve the efficiency of decisions.} Moreover,  optimal mechanisms may involve random score submission  recommendations. It is also possible that the same submitted score is assigned to a different distribution over final decisions depending on the type report that triggered it.  For example, consider an agent whose natural score is their musical talent and soft information is their privately-known tastes for extracurricular activities. Depending on the designer's preferences the mapping from submitted music performance recordings to decisions (e.g., level of aid, major etc.) can vary with a student's type report and depend on student tastes.\footnote{See the example in \secref{sec-ex-football} for an illustrative story.} Such mechanisms rely on a mediator and on reports from the agent.

In \secref{sec-score-based} we provide conditions under which conditioning the score-based decision rule on soft information is not needed and thus scoring-based mechanisms (that only condition on scores) are without loss of generality. In other words, we identify situations under which the agent does not make any type reports and decides without a mediator what score to submit in a score-based decision rule. A key condition is that the original mechanism does not rely on random score submission requests. This is de facto the case in settings in which the designer wants to ensure that agents submit their natural scores and restricts attention to falsification-proof mechanisms  (see \cite{perez2023fraud} for example).

In \secref{sec-bin-opt} we characterize optimal mechanisms to screen a persuader in the presence of manipulable hard information. Put differently, we design optimal ordeal mechanisms or costly screening mechanisms. We do so using a methodology that is suitable  both for concave cost functions and quadratic costs. In several leading settings the optimal mechanisms we characterize are implementable via score-based rules and do not even require commitment to the decision which can be taken by a third party, a decision maker that is a stand-in for consumers, firms and so forth.  This new work provides a framework and methodologies  applicable in a wide range of contexts in economics and in computer science.

\subsection{Related literature}

\paragraph{Literature on mechanism design with evidence} 
We contribute to the literature on mechanism design with evidence and in particular to that with moderate misreporting costs by providing a formulation that allows both for soft and hard information and by showing how this setting can be casted as a generalized principal-agent setting. Leveraging the revelation principle, we show how canonical mechanisms decompose into two mappings:  a \textit{score recommendation rule} and, a possibly, type-dependent \textit{score-based decision rule}. The independent work by \cite{strausz2022principled} is in the same spirit but analyzes a setting with $0$ or $\infty$ costs and in which the implementable outcomes do not include the payoff relevant implications of presenting evidence.

\textit{Infinite evidence costs:} In the classical formulation of mechanism design settings with evidence, an agent either has or does not have a piece of evidence (0 or infinity cost)
and evidence is submitted as an input message in the mechanism see \cite{green1986partially,forges2005communication,bull2007hard,deneckere2008mechanism,ben2012implementation}. \cite{green1986partially} are the first to note that the revelation principle fails in the sense that some social choice functions can only be implemented with partial evidence and provide conditions on the evidence structure, called nested range condition,  under which  the set of implementable social choice functions coincides with the set of truthfully implementable social choice functions. The subsequent papers \cite{forges2005communication,bull2007hard,deneckere2008mechanism} provide alternative conditions on the evidence structure available to agents (normality) such that presenting \emph{maximal} evidence is without loss of generality recovering, in this weaker sense, the revelation principle. The reasons why truth-telling (in an appropriate sense) fails in settings with evidence, is simple: it not only matters which type(s) $t$ can mimic (call them $\mathcal{T}(t)$), but also which types in $\mathcal{T}(t)$ can mimic. Normality and the related conditions, guarantee that if $t$ can mimic $t'$, $t$ can also mimic any type $t'$ can mimic.

\textit{Moderate evidence costs:} \cite{bull2008costly} studies costly evidence production  by two agents in a court setting and analyzes its effect of court outcomes when settlement and non-settlement are possible. \cite{bull2008mechanism} allows for moderate linear evidence costs and  shows that the sufficiency of the special three-stage mechanism of \cite{bull2007hard}  holds also with moderate evidence cost. The mechanisms in \cite{bull2008mechanism}  differ from those in \cite{bull2007hard}  in that they allow for a public signal from the external enforcer. Since transfers can be used to motivate the disclosure of evidence in the third stage, the second-stage signal can be public. However, transfers do not eliminate the need for the second stage because randomization by the external enforcer may be needed. \cite{kartik2012implementation} study  Nash implementation  (as in \citealp{maskin1999nash}) with evidence and their setting nests costly and hard evidence. By contrast, we  show how mechanism design with evidence, regardless of the cost structure, can be cast with the original formulation of \cite{myerson1982optimal}.

\paragraph{Literature on mechanism design with costly misreporting} 
We also contribute to the literature on mechanism design with costly misreporting (\citet{lac_wei89,maggi1995costly,crocker1998honesty}) by showing that indeed the revelation principle in \cite{myerson1982optimal} applies and by underlining the parts of the mechanism where randomization is needed.  \citet{lac_wei89} incorporate costly state falsification in a model of risk-sharing contracts and characterize optimal falsification-proof contracts, but also show they may be outperformed by contracts that induce falsification. \cite{maggi1995costly}, \cite{crocker1998honesty} derive mechanisms in settings with  costly state falsification in single-agent settings with transfers. In 
\cite{crocker1998honesty} the contract specifies transfers and an action to be taken by the agent as a function of his type report $x$, denoted by $y(x)$. In some interpretations of their abstract model, the distance $|y-x|$ affects falsification costs. As is the case in \cite{maggi1995costly}, the optimal contract in \cite{crocker1998honesty} relies on distortions on $x$. \cite{severinov2019screening} focus on a mechanisms with transfers and  provide conditions on reporting costs to ensure truth-telling is without loss. \cite{deneckere2022signalling} show that in environments with misrepresentation costs having agents send multiple signals, significantly expands the set of implementable outcomes and results to near efficiency when misrepresentation costs are small. \cite{tan2023price} considers a price discrimination setting in which agents engage in costly behavior distortions to avoid being discriminated. 
By contrast, there are no transfers in \cite{perez2022test} who derive optimal tests in an agent-decision maker setting (a.k.a sender-receiver setting) in which the agent can falsify at a cost inputs into the test. Finally, there is also a computer science literature on mechanism design with reporting costs, most notably \citep{kephart2016revelation} who provide conditions on reporting costs to ensure truth-telling is without loss.

\paragraph{Literature on costly signaling and screening.} We also relate to the vast literature stemming from \cite{spence1978job}'s classic signaling model and the recent literature that studies costly gaming distortions  in signaling settings stemming from the important contribution of \cite{frankel2019muddled}. These works include \cite{ball2020scoring} and \cite{frankel2020improving}. We differ in that we consider general mechanisms (rather than linear scoring rules). Our characterization of optimal approval mechanisms in settings without transfers relates to the literature on costly screening via various tools: 
notably money burning as studied, among others, in \cite{hartline2008optimal}, \cite{condorelli2012money}, \cite{chakravarty2013optimal}. In those settings the utility burnt does not depend on the agent's type whereas in our setting falsification costs are type-dependent. 
 \cite{dworczak2022equity} studies costly ordeals in a specialized setting with linear costs and allowing for deterministic mechanisms. In  \cite{dworczak2022equity}  the cost to achieve ordeal $y$ is linear in type.  The allocation is an amount of money $x \in \mathbb{R}$ whereas we have an approval probability that must be in $[0,1]$. This is not a crucial difference.  In  \cite{dworczak2022equity} and in most papers studying ordeals (costly screening mechanisms) the cost to achieve a given level of benefit is increasing in type. In our setting, each type has a different ordeal level that costs zero (this is the ordeal level that corresponds to presenting the natural score). The optimal mechanisms we design simultaneously specify how each submitted score is mapped to an approval probability and a (random) score submission recommendation rule as a function of reported types. By contrast to these earlier works, the mechanisms we allow for can depend jointly on the ordeal and the type performing the given ordeal. This feature can be important for applications where the designer wants the allocation to depend jointly on the task and the type of agents because certain types have higher weight on designer's objective function.\footnote{ \cite{akbarpour2023comparison} analyze how to rank various costly screening tools. By contrast, we study given exogenously given falsification costs, how to optimally screen agents.}
 
  Our characterization of optimal approval mechanisms in settings without transfers also relates to \cite{li2024screening} who study costly screening in a multi-agent setting without transfers and identify conditions under which contests are optimal and situations under which random mechanisms dominate contests. 
We differ in the cost structures we examine and the objective function.\footnote{As discussed in \cite{li2024screening}, mechanism design in the presence of costly manipulations, relates to works in computer science that study strategic classification. See, for instance, \cite{hardt2016strategic} who present an efficient classification algorithm that minimizes errors in the presence of gaming.}

\paragraph{Literature on persuasion mechanisms with evidence}
\cite{glazer2004optimal, gla_rub06} study settings in which the sender seeks to persuade the receiver to `accept' them regardless of the state of world whereas whether the receiver prefers to accept or to reject depends on the state of the world. The sender knows the state of the world and can present the receiver hard evidence about it. They derive optimal persuasion rules. These rules maximize the probability that the listener accepts the request if and only if it is justified. In \cite{gla_rub06} the authors show that neither commitment to the decision nor randomizations have any value. In \cite{glazer2004optimal} the sender can, in addition, send an arbitrary (cheap talk) message to the receiver and upon receiving the message the receiver can request hard evidence. Like in our example  in \secref{sec-ex-football}, it is often beneficial for the receiver to randomise in the evidence she asks for from the sender. However, by contrast to our example  in \secref{sec-ex-football}, in \cite{glazer2004optimal} it is not beneficial to randomise the final decision of accepting or rejecting once the evidence has been provided. \cite{sher2011credibility} provides generalizations to the aforementioned results and identifies the key conditions on payoffs (namely, concavity) that renders commitment to have no value. \cite*{hart2017evidence} focus on truth-leaning equilibrium and identify the structure of evidence that guarantees that commitment cannot yield any advantage.\footnote{Committing to a mechanism also has no value in the allocation setting without transfers considered by \cite{ben2019mechanisms}.}

\section{Model}\label{sec-model}

We consider an uninformed principal facing a privately-informed agent.\footnote{We can straightforwardly extend the setting to accommodate multiple agents at the expense of more cumbersome notation.} The agent's type, denoted as $\type=(\theta,\sn) \in \Types$, consists of two components. First, $\theta \in \Theta$ represents the agent's preferences, abilities, and needs, which is soft information and can be misrepresented without cost. Second, $\sn \in \Sn$ stands for the agent's natural score or evidence, such as a transcript or a certificate of disability, which is costly to falsify. There is a commonly-known prior over $\Types$, denoted by $F \in \Delta(\Types)$. The falsification cost function $c: \Ac \times \Types \to \mathbb{R}_+$ defines the cost for a type $\type=(\theta,\sn)$ to present a score $\ac \in \Ac$, with $\Ac \subseteq \Sn$. We assume that presenting the true score is costless, i.e., $c(\sn,\type)=0; \forall \type=(\theta,\sn)$. It is important to note that the cost of \ac\ depends not only on the natural score $\sn$ but also on $\theta$, capturing aspects like gaming ability or discomfort from lying.

We denote by $X$ the set outcomes or decisions. An outcome $x$
 can stand for quality-transfer pairs as in \cite{mus_ros78}, levels of aid, bonuses, promotions and so forth. The agent's payoff is a function $u: X \times \Ac \times \Types \to \mathbb{R}$, defined as $$u(x,\ac,\type)= v(x,\type) - c(\ac,\type).$$ Let $\Fals$ denote the set of functions $\fals:\Types \to \Delta(\Ac)$. In what follows, $\fals$ stands for the agent's falsification strategy. 
The designer's payoff is a function 
$u: X \times \Ac \times \Types \times \Fals \to \mathbb{R}$. Note that the designer's payoff does not only depend on the joint distribution of decisions and types (that is $X\times T$), rather it depends on the joint distribution over decisions, types, scores submitted as well as the agent's falsification strategy. It can be of the form $$u_P(x,\ac,\type,\rho)= v_P(x,\type,\ac) - c_P(\fals,\type),$$ capturing that the designer may internalize the agent's burnt utility caused from falsification. All the sets $\Types, \Ac, \Sn,\Theta$ are finite.

  \section{Canonical mechanisms}\label{sec-can-mech}
In this section we show how to cast our setting to a generalized principal-agent setting in \cite{myerson1982optimal} and describe the canonical class of mechanisms. 

To do so, we treat a score submission both as a type report and as a payoff-relevant action. An agent submits a type report which contains an informal report about their natural score, which is costless no matter the score they claim to have, and a formal submission of a score, which can be costly if they falsify their natural score. The formal score submission is payoff-relevant and analogous to an action in \cite{myerson1982optimal}. The revelation principle in  \cite{myerson1982optimal} establishes that any outcome, in other words, any joint distribution on $X \times \Ac  \times \Types$, arising at a BNE of any abstract mechanism, arises at a truthful and obedient equilibrium of a direct recommendation mechanism:   
A DRM maps type reports to a distribution over contractible outcomes (denoted by $D_0$ in \citealp{myerson1982optimal}) and action recommendations, one for each player. 

The formal submission of a score $\ac \in \Ac$ corresponds to choosing an action in \cite{myerson1982optimal}. The contractible outcomes in our setting, denoted by $\Cd$, are functions from observable score submissions (actions) to distributions over decisions: 
\begin{align*}
\Cd= \{ \aloc: \Ac \to \Delta(X) \}.
\end{align*}
With our notation direct recommendation mechanisms are defined as follows:
\begin{align*}
 & \mu: \underbrace{\Types }_{ \Theta \times \Sn} \to \Delta(\Cd \times \Ac)  \iff \mu: \underbrace{\Types}_{ \Theta \times \Sn} \to \Delta\left(\Delta(X)^\Ac \times \Ac\right).
\end{align*}  
Observe that each type report \type\ triggers a distribution over score requests ($\ac \in \Ac$) and a distribution over score-based decision rules ($\aloc \in \Cd$). The mechanism then determines the ultimate joint distribution over types, decisions and submitted scores which matter because they are payoff relevant. Observe also that the mechanism may request falsified scores.  


To summarize, in this formulation the agent reports a type $\type=(\theta,\sn)$ and presents evidence $\ac$. A piece of evidence or \textit{score} is submitted twice: first, informally via a costless message  as part of the type report and second, as a formal submission which can be costly and amounts to choosing a payoff-relevant action in \cite{myerson1982optimal}.  There are several real-world situations that resemble this procedure. In a job application setting, the costless message is to state a university degree on a CV and the formal submission is to show the certificate.
In a disaster relief application, the costless message is the claim there was flood damage and the costly action is to produce evidence of the damage. Similarly, in a school choice setting the costless message amounts to stating the address in the form and the costly action amounts to producing evidence of this address (a utility bill, lease or ownership contract).  

As part of the type report, the agent can freely lie about the score but the formal submission can be costly when the natural score is falsified. 

 \begin{Prop}[Revelation principle](\citealp{myerson1982optimal}) Any BNE equilibrium feasible joint distribution on $T\times X \times S$ arising at a BNE of any abstract mechanism, arises at a truthfull and obedient equilibrium of a direct recommendation mechanism.
 \end{Prop} 
 \textit{Sketch of proof}
 As a reminder, to obtain the result we start with some indirect mechanism $\pi: \mathcal{R} \to \Delta(D \times \mathcal{M})$  where $\mathcal{R}$ ($\mathcal{M}$) stand for  abstract input (output) messages. The agent employs a reporting rule $\sigma: \Types \to \Delta(\mathcal{R})$ and chooses an action $\ac \in \Ac$ as a function of message $m \in \mathcal{M}$ received using action rule $\delta: \mathcal{M} \to \Delta(\Ac)$. The agent's strategy consists of the reporting and the action rule which together with the indirect mechanism $\pi$ determine the outcome which is a joint distribution over $X\times \Types \times \Ac$. Alternatively, the mechanism $\pi$ and the agent's reporting and action rule can be thought of as transition probabilities\footnote{When we refer to a mapping as transition probability from one set to another we remove the $\Delta$.}
 \[ \sigma: \Types \to \mathcal{R}; \pi: \mathcal{R} \to D \times \mathcal{M}; \delta: \mathcal{M} \to \Ac\]
 and the corresponding DRM is a transition probability built from composing $\pi,\sigma,\delta$ as follows: 
  \[ \mu= \sigma \circ \pi \circ \delta : \Types \to D \times \Ac. \]
  It is straightforward to argue that if $(\sigma,\delta)$ are part of a Bayes-Nash equilibrium given the indirect mechanism $\pi$, truth-telling and obedience are optimal for the agent given the DRM $\mu$. Mechanism $\mu$ by construction results to the same joint distribution over $X \times \Types \times \Ac$ and to the same transition probability $\fals: \Types \to \Ac$  as $\pi,\sigma,\delta$ do. 
 
Thus the proof of \cite{myerson1982optimal} straightforwardly extends to the proposed setting. The fact that the designer's payoff can depend on  the $\fals: \Types \to \Delta(A)$ is immaterial for the argument. 

\paragraph{Incentive-compatible DRMs}Incentive compatibility constraints split into truth-telling and obedience constraints:
%

\small
\begin{align}\label{eq:IC-obs-3}
&\hspace{-0.4in}\sum_{\aloc \in \Cd } \sum_{\ac\in \Ac}\mu (\aloc,\ac \mid \type) \left[\sum_{x\in X}\aloc(x|\ac)v(x,\type)-c(\ac,\type)\right]  \geq \sum_{\aloc \in \Cd }\sum_{\ac\in \Ac}\mu (\aloc,\ac \mid \typeb) \left[\sum_{x\in X}\aloc(x|\ac)v(x,\type)-c(\ac,\type)\right] \quad \forall\ \type,\typeb \in \Types \tag{TT} \\
&\hspace{-0.4in}\sum_{\aloc \in \Cd }\mu (\aloc,\ac \mid \type) \left[\sum_{x\in X}\aloc(x|\ac)v(x,\type)-c(\ac,\type)\right]  \geq \sum_{\aloc \in \Cd }\mu (\aloc,\ac \mid \type)  \left[\sum_{x\in X}\aloc(x|\acb)v(x,\type)-c(\acb,\type)\right]  \forall\, \type \in \Types; \ac \in \supp \mu(\cdot |\type), \acb \in \Ac.  \tag{OB}
\end{align}
\normalsize

If the agent's participation in the mechanism is voluntary then the mechanism must also satisfy participation constraints:
\begin{align}\label{eq:PC}
 \sum_{\aloc \in \Cd }\mu (\aloc,\ac \mid \type) \left[\sum_{x\in X}\aloc(x|\ac)v(x,\type)-c(\ac,\type)\right]  \geq \underline{u}(\type) \quad\  \forall\, \type\in \Types,  \ac \in \supp \mu(\cdot |t)  \tag{PC}
\end{align}
where $\underline{u}(\type)$ denotes the agent's payoff from non-participation. 
For future reference we also let:
\begin{align}\label{eq:lie-payoff}
U(\type,\typeb,\mu) \equiv \sum_{\aloc \in \Cd }\sum_{\ac\in \Ac} \mu (\aloc,\ac \mid \typeb) \left[\sum_{x\in X}\aloc(x|\ac)v(x,\type)-c(\ac,\type)\right]. \end{align}

Earlier works on mechanism design settings with costly misrepresentation of types e.g. \citep{kephart2016revelation,severinov2019screening} treated the score submission analogous to a type report and identified conditions under which submitting the natural score is without loss of generality. With this more traditional view, the revelation principle (which is often used as a synonym with truth-telling) could fail. Indeed, as discussed in the introduction, in settings without monetary transfers such as the ones considered in \cite{perez2022test} restricting attention to mechanisms that incentivize the agent to submit the natural score are with loss of optimality. This is true even for falsification cost functions that satisfy the triangular inequality considered in \cite{kephart2016revelation} among others. Instead, here by treating the score submission as a payoff-relevant action we recover the generalized revelation principle by  \cite{myerson1982optimal}. However, now, while the agent is reporting truthfully, the score submission request rule may ask the agent to falsify and it is not easy to identify a priori the optimal falsification targets for each type. We proceed to obtain further simplifications that we later leverage to solve for optimal mechanisms.

\begin{Lem}[Wlog deterministic mechanisms over score-based decision rules]\label{lem-no-random-score}Within our formulation, there is no need to randomize over score-based decision rules: each type report triggers a single $\aloc \in \Cd$. 
\end{Lem}

The intuition behind \lemmaref{lem-no-random-score} is simple: Given that a score-based decision rule $\aloc: \Ac \to \Delta(X)$, specifies a randomization over $x$'s we do not need randomization over $\aloc$'s: Conditional on a type report $t$
the mechanism specifies a unique score-based decision rule but possibly to random score submission requests.

\paragraph{Mechanism decomposition} In light of \lemmaref{lem-no-random-score} mechanisms simplify from $\mu: \Types \to \Delta (\Cd \times \Ac)$ to
$ \mu: \Types \to \Cd \times\Delta(\Ac)$
and, more importantly, a mechanism $\mu$ decomposes into two mappings: A \textit{score-based decision rule:}  \[\alocm: \Ac \times \Types \to \Delta(X)\] defined from $\mu$ as follows $\aloc(x | \ac,\type)\equiv \aloc(x|\ac)\mathbb{1}_{\mu(\cdot|\type)=\delta_{\aloc}}$, and a \textit{score recommendation rule:} \[\fals: \Types \to \Delta (\Ac).\]
The mapping $\fals$ plays the role of the agent's falsification strategy and maps a type report to  random score submission requests.

\paragraph{Voluntary participation}
To formally accommodate the agent's participation decision we add additional actions in \Ac. In what follows, we abuse notation and take \Ac\ to contain the scores that the agent submits and, in addition, the decisions to participate or not to participate. 

In our setting, score submission is observable and hence, the mechanism can assign the null outcome or non-participation outcome (call it $\underline{x}$), if the agent inputs a score $\acb$ in $\aloc$ instead of the score $\ac$ requested by $\mu$. With this observation, the obedience constraints boil down to ex-post participation constraints. Here ex-post means conditional on a score submission recommendation rather than when we sum all recommendation on the support of \fals. We summarize this observation in the following lemma:

 \begin{Lem}[Obedience implied by voluntary participation ]\label{lem-PC-to-OB} Let $\underline{x}$ denote the non-participation outcome and let $\underline{u}(\type)$ denote the corresponding payoff of type $t$. Within our formulation, obedience constraints reduce to participation constraints. 
\end{Lem}
Note that in the presence of a non-participation null outcome, the agent has to obey score submission requests that he is potentially not indifferent among.
\paragraph{Simplified mechanisms} From the above, it follows that it is without loss to focus on direct recommendation mechanisms with allocation rule
$\alocm: \Ac \times \Types \to \Delta(X)$ and score recommendation rule $\fals: \Types \to \Delta (\Ac)$ that satisfy \textit{obedience:}
\[\mathbb{E}_{\alocm(\ac,\type)}v(x,\type)-c(\ac,\type)\geq 0, ~~~~~~~\forall a\in\supp \fals(\type)\]
and \textit{truth-telling}:  
\[U(\type)\equiv\mathbb{E}_{\fals(\type)}\left(\mathbb{E}_{\alocm(\ac,\type)}v(x,\type)-c(\ac,\type)\right)\geq \mathbb{E}_{\fals(\typeb)}\left\{\mathbb{E}_{\alocm(\ac,\typeb)}v(x,\type)-c(\ac,\type)\right\}^+,~~~~~~~\forall \type,\typeb.\]


\section{Score-based mechanisms}\label{sec-score-based}

 A \textit{score-based mechanism} is an allocation rule based only on (final) score $\ac$, $\aloc:\Ac\to\Delta (X)$. A \textit{falsification strategy} $\sig:T\to\Delta (A)$ is incentive compatible if
\[\mathbb{E}_{\aloc(\ac)}v(x,\type)-c(\ac,\type)\geq \mathbb{E}_{\aloc(\acb)}v(x,\type)-c(\acb,\type),~~~~\forall \ac\in \supp \sig(\type),~\acb\in \Ac.\]

We proceed to provide sufficient conditions for a score-based mechanism to be without loss of generality. 

\begin{Ass}[Separability assumptions]Suppose the \textit{agent's preferences} can be written as \[u_A(x,\ac,\type)=\beta(\ac,\type)v(x)-c(\ac,\type)\] with $\beta(\ac,\type)>0$ and the \textit{principal's preferences} are:
\[u_P(x,\ac,\type)=w(\ac,\type)r(x,\ac)+y(\ac,\type)v(x)+\ell(\ac,\type)\] with $w(\ac,\type)>0$.
\end{Ass}
 Note that the assumptions are automatically satisfied if $|X|=2$ by normalization.

\begin{Prop}[Score-based principle]\label{prop-score}
If preferences satisfy the \textit{separability assumptions} and $(\alocm,\fals)$ is an incentive compatible direct mechanism with a \textit{deterministic} score recommendation rule, then there exists a \textit{score-based mechanism} $\aloc$ with an incentive compatible and \textit{falsification strategy} $\sig$ such that maintains the agent's payoff and weakly increases the principal's payoff.
\end{Prop}

The result in \propref{prop-score} is analogous to the \textit{taxation principle} whereby instead of submitting a report in a mechanism, the agent chooses an option from a menu. Here the agent only decides what score to submit in a fixed score-based decision rule, $\alocs: \Ac \to \Delta(X)$. Instead in a mechanism, the allocation rule can depend on the agent's type as well, so $\alocm: \Ac\times\Types \to \Delta(X)$. In other words, in a mechanism two different types \type\ and \typeb\ can be submitting the same score but face different distributions over decisions. When the mechanism relies on a deterministic score recommendation rule, incentive-compatibility implies that types  \type\ and \typeb\ are indifferent between each others distributions over decisions. 
The designer separability condition is necessary for the designer to always prefer one stochastic allocation for both types whenever two agent types submitting the same score are indifferent across two distributions over decisions. Principal separability is required because a stochastic allocation is multidimensional unlike a transfer (and if  $|X|=2$, no condition on the designer's payoffs is required).
Agent separability plays an analogous role as the standard quasilinearity assumption. 
\paragraph{Implications}
Suppose that the designer wants to restrict attention to falsification-proof mechanisms as is done in \cite{perez2023fraud}\footnote{Falsification-proof mechanisms do not burden agents and avoid negative externalities. In the words of \cite{pathak2008leveling}, falsification-proof mechanisms level the playing field.}  then \propref{prop-score} implies that restricting attention to score-based mechanisms is without loss of generality. 

\begin{Def}[Falsification proof mechanisms] An IC mechanism is falsification proof if it is a best response for all types $t=(\theta,\sn) \in T$ submit their true natural score, that is $\ac=\sn$. 
\end{Def}
\begin{Cor}\label{cor-score}In a multi-outcome generalized persuasion setting any IC falsification proof mechanism  can be replicated by a scoring mechanism. Thus, the mapping from submitted scores to decisions cannot depend on soft dimensions of types. 
\end{Cor}

\begin{Rem}[Comparison: tests versus score-based decision rules versus mechanisms] To understand the differences between general mechanisms versus score-based mechanisms considered in \cite{perez2023fraud} (which as stated in \corref{cor-score} are without loss in that setting) versus tests considered in \cite{perez2022test} we now compare them in our setting. 

\textit{Test \& falsification:} Faced with a test $\tau: \Ac \to \Delta(\mathcal{M})$
the  agent chooses a  falsification strategy that maps a natural  score to a falsified score. The test converts the possibly falsified score to a signal $m \in \mathcal{M}$. There is a third party, a decision maker, who upon observing the signal $m$, makes a decision $X$. 

\textit{Score-based mechanism \& falsification:} Now there is a score-based mechanism  $\aloc: \Ac \to \Delta(X)$. Faced with such a mechanism 
the  agent chooses a  falsification strategy that maps a natural  score to a falsified score. The mechanism maps a submitted score to a decision.  There is commitment to the decision and no separate decision-maker. 

\textit{General mechanisms} A general mechanism is \begin{align*}
\aloc: \Types\times \Ac \to \Delta(X) \text{ and } \fals: \Types \to \Delta(\Ac)
\end{align*}
The agent submits a type report and the mechanism maps the  report to a score-based decision rule and a score submission request. The revelation principle tells us that truth-telling and obedience are without loss. The difference between score based and general mechanisms is that in general $\aloc$ can vary with soft dimensions of type in contrast to score-based mechanisms. 
\end{Rem}
We illustrate these differences in a simple example in what follows. 

\subsection{Illustrative example: college admission}\label{sec-ex-football}

The designer is a college facing a student with four equally likely types:
$$T=\{ (F,\sn_L), (NF,\sn_L), (F,\sn_H), (NF,\sn_H)\}.$$ The first element of each type describes whether or not the student likes football (F or NF) whereas the second element is the natural score which can be low $(\sn_L)$ or high $(\sn_H)$. The decision is binary, so $X=\{0,1\}$, where $0$ stands for not admit while $1$ stands for admit. 
The cost to falsify to $\sn_j \neq \sn_i$ is $1$ for all types. The payoffs from each decision as a function of the agent's type are summarized in the table below where the first number is the agent's payoff whereas the second number is the designer's payoff: 

\begin{center}
\begin{tabular}{|c|c|c|}
\hline
		&  $1$  		&$0$  	    \\ \hline  
\small $\type_1=(F,\sn_L) $  	&   $1, 3$	   					& $0,0$        \\ \hline
\small $\type_2=(NF,\sn_L)$ 	&  $1,-1$  				       & $0,0$    \\ \hline

\small $\type_3=(NF,\sn_H)$ 	& $1,2$    				       &   $0,0$ \\ \hline
\small $\type_4=(F,\sn_H)$ 	& $1,4$   				       & $0,0$ \\
\hline 
\end{tabular}
\captionof{table}{Student and college payoffs}\label{table-payoffs}
\end{center}

\underline{  Scenario 1: } Suppose that the designer's payoff depends only on the admission decision. The designer's (here, the college's) first-best is to admit everyone except $\type_2$. This can be achieved by 
faces 
the following test:
\begin{align*}
\aloc(\sn_L)=0, \aloc(\sn_H)=1 
\end{align*}
$\type_2,\type_3,\type_4$ do not falsify; $\type_1$ falsifies $\sn_L$ to $\sn_H$ and thus gets admitted but burns all utility because the cost of falsification is $1$.
\underline{  Scenario 2: } In this scenario, the designer's payoff depends on the decision and the incurred falsification costs. In particular,  the loss function is quadratic incurred costs $c$   $L(c)=\frac{c^2}{6}$. Table \ref{table-payoffs} lists the payoffs corresponding to each decision. To get the total payoff for the designer we subtract the loss due to falsification cost.   Consider the following menu of score-based decision rules:
\begin{align*}
&\text{ pooling at $0$: } \aloc^0(\sn_L)=\aloc^0(\sn_H)=0 \\
&\text{ pooling at $1$: } \aloc^1(\sn_L)=\aloc^1(\sn_H)=1 \\
&\text{ separating: }\aloc^S(\sn_L)=0, \aloc^S(\sn_H)=1 \\
&\text{partially separating: }\aloc^{PS}(\sn_L)=\frac14, \aloc^{PS}(\sn_H)=1 .
\end{align*}
The optimal assignment to a score-based decision rule of each type is as follows:
\begin{align*} 
& \alpha(\aloc^1|\type_1)=1  \\
& \alpha(\aloc^{PS}|\type_2)=1 \\
& \alpha(\aloc^S|\type_3)=1 \\
& \alpha(\aloc^S|\type_4)=1.
\end{align*}
Finally, the optimal score submission request rule is:
\begin{align*}
&\fals(\sn_L|\type_1)=\frac{1}{4}, \fals(\sn_H|\type_1)=\frac{3}{4}\\
&\fals(\sn_L|\type_2)=1, \fals(\sn_H|\type_2)=0 \\
&\fals(\sn_L|\type_3)=0, \fals(\sn_H|\type_3)=1 \\
&\fals(\sn_L|\type_4)=0, \fals(\sn_H|\type_4)=1.
\end{align*}
Note that this mechanism satisfies truth-telling and obedience. Types $\type_1,\type_2$ get $\frac14$ if they mimic each other and get zero if they mimic $\type_3$ or $\type_4$. Types $\type_3,\type_4$ get their maximum payoffs by reporting the truth. 
Whereas in Scenario 1 the optimum can be achieved by a test that entails no communication nor commitment on the decisions; in scenario 2 the optimum needs communication and commitment: it is a  mechanism and not a test. In scenario 2, the designer's optimal mechanism \textit{requests scores stochastically} and different type reports lead to \textit{different} score-based rule. By contrast to the findings in \cite{glazer2004optimal}, \cite{sher2011credibility} and \cite{hart2017evidence} \cite{ben2019mechanisms}  
we have a pure persuasion setting with binary actions in which (i) commitment is valuable, (ii) communication is valuable, (iii) randomization in evidence requests is valuable and (iv) randomization in decisions is valuable. The difference with the earlier papers lies in that we consider general mechanisms that allow for randomizations and in that information in our setting is semi-hard and thus falsification is payoff-relevant. This example shows that scoring mechanisms (which, by contrast to tests encode commitment to the decision) can be dominated by a mechanism that receives type reports from the agent and outputs random score submission recommendations.
 
\section{Optimally screening a persuader: binary outcomes}\label{sec-bin-opt}

In this section we derive optimal mechanisms for the designer in a binary outcome setting in which $X=\{0,1\}$ and agent types equal natural scores  $\Types=[\smin,\smax]$, with $\smin<0<\smax$ so $\type=\sn$ and there are no soft dimensions of the type. The distribution of types $F$ has full support and strictly positive density $f$. To fix ideas, we call decision $1$ approval and decision $0$ rejection. Regardless of type, the agent is a persuader who wants to be approved, so prefers $x=1$ to $0$, whereas the designer wants to approve only positive types. There are no transfers. This binary outcome setting captures many leading settings such as allocation of a good to an agent with unit demand without transfers (the agent either gets a unit or not); acceptance decisions, approvals, promotions and many other settings. As discussed in the introduction, analogous settings have been analyzed in \cite{glazer2004optimal}, \cite{sher2011credibility} and \cite{perez2023fraud}. \cite{li2024screening} study a richer allocation problem with many goods and agents under linear signaling costs.\footnote{Under certain conditions, their findings relating to whether or not contests are dominated by mechanisms that involve randomness go through under convex costs as they explore in their appendix.}  

The agent, regardless of type, gets a payoff of $1$ if approved and $0$ otherwise, that is for all $\type \in \Types$:
\[ u(x,\ac,\type)= v(x,\type) - c(\ac,\type)=\begin{cases}1 - c(\ac,\type) \text{ if } x=1\\
- c(\ac,\type) \text{ if } x=0 \end{cases}. \]
The designer's outside option from rejecting the agent is $0$. The designer's payoff from accepting an agent of type \type\ is equal to  \type, therefore the first-best is to accept positive types and to reject negative ones. There are no resource constraints. The cost function is scaled by $\gamma >0$ which stands for the agent's gaming ability assumed to be known, so  $c: \Types \times \Ac  \to \mathbb{R}$ and $\cscale c(\ac,\type)$ denotes the cost to type \type\ of submitting score \ac. Types are distributed according to a commonly-known full support distribution $F$. We focus on the case that $\mathbb{E}_{F}[\type]<0$. We assume that no falsification is costless so when $\ac=\type$, $\cscale c(\ac,\type)=0$ (which is just a normalization) and that for all $\ac\geq\type$ the cost is decreasing in \type.

For this binary outcome setting, the allocation rule simplifies to an approval probability as a function of a type report \type\ and a submitted score \ac : 
\[\aloc(\ac,\type)\in[0,1].\]
The mechanism, therefore, consists of an allocation rule that depends both on the agent's type \type\ and submitted score \ac, $\aloc: \Ac \times \Types \to [0,1]$ and score recommendation rule $\fals: \Types \to \Delta (\Ac)$.

The interim approval probability is the expectation over all recommended scores: 
\begin{align}\label{eq-u-agent}
Q(\type)=\int_A\aloc(\ac,\type)d\fals(\ac,\type).
\end{align}
The agent's payoff simplifies to  $U(\type) =Q(\type) -\int_{\ac\in \Ac}\csa d\fals(\ac| \type)$ and the ex-post participation constraints (which, as mentioned earlier, imply obedience) and truth-telling constraints write:

\begin{align*}
 &\aloc(\ac,\type) -\csa\geq 0 \quad \forall \ac \in \supp  \fals(\cdot | \type) \tag{PC}\\
 &U(\type)\geq \int_{\ac\in \Ac}  \left[ \aloc(\ac,\typeb) -\csa\right]d\fals(\ac| \typeb)= Q(\typeb)- \int_{\ac\in \Ac}   \csa d\fals(\ac|\typeb)  \quad \forall \type,\typeb \in \Types. \tag{TT}
\end{align*}

\paragraph{Designer's objective}
The designer seeks the mechanism that solves
\begin{align*}
&\max_{\aloc,\fals } \int_{\smin}^{\smax}Q(\type)\type f(\type)d \type \\
&\text{ subject to TT, PC, probability constraints}. 
\end{align*}
In what follows, when a  score recommendation rule $\fals$ is deterministic for each \type\ (a Dirac on some \ac) we denote it simply as $\ac^*: \Types \to \Ac$. 

\textbf{First best} The first-best for the designer is:
\begin{align}
Q^{\text{FB}}(\type)&=\begin{cases} 0 \text{ for } \type < 0\\
1\text{ for } \type\geq 0.
\end{cases}
\end{align}
If $1-\cscale c(\smax,0) <0$, so $\gamma <c(\smax,0)$ we can achieve the first best by setting \begin{align*}
\ac^*( \type)= \begin{cases} \type\text{ for }\type \in [ \type^{\star},\smax]\\
\type^{\star} \text{ for }\type \in [0, \type^{\star})\\
\type \text{ for }\type<0  \end{cases} \quad \aloc( \ac, \type)=\begin{cases} 1 \text{ for }\ac \geq \type^{\star}\\
0 \text{ for } \ac \neq \type^{\star}  \end{cases}
\end{align*}
where $\type^{\star}$ satisfies $1-\cscale c( \type^{\star},0)=0$. In words,  $\type^{\star}$ is the highest score type $0$ is willing to falsify to in order to get approved with probability $1$. All positive types get approved with probability 1. Positive types up to $\type^{\star}$ falsify to $\type^{\star}$ while all positive types above $\type^{\star}$ do not falsify. All negatives get approved with probability $0$ and do not falsify.\footnote{The definition of $\type^{\star}$ and the fact that $c$ is decreasing in $\type$ imply together that $1-\cscale c( \type^{\star},\type)<0$ for $\type<0$.} 

In what follows, we solve for the optimal mechanism when $1-\cscale c(\smax,0)\geq 0$ so 
\begin{align}\label{eq:gamma-ob}
\gamma > c(\smax,0) 
\end{align} holds and therefore falsification costs are low enough to make the first-best impossible.

\begin{Lem}\label{LemMonAloc}Without loss of optimality we can restrict attention to $\aloc: \Ac \times \Types \to [0,1]$  increasing in $\ac$ for all $\type$. 
\end{Lem}

\subsection{Concave costs}
 Suppose that for each $\ac \in \Ac$, with $\ac\geq \type$, $c(\cdot, \type)$ is decreasing and concave in $\type$. Recall that in this section, $\type \in \Types=[ \smin, \smax ]\subset \mathbb{R}$. 
\paragraph{Implications of incentive compatibility}
The agent's payoff from truth-telling writes:
\[ U(\type)=\max_{\typeb \in \Types} \left(Q(\typeb)- \int_{\ac\in \Ac}\csa d\fals(\ac| \typeb) \right)\]
and when $c$ is concave it is convex as it is the maximum of convex functions. 
Let 
\begin{align}\label{eq-int-rep-0-new}
C(\type)\equiv - \int_{\ac\in \Ac}  \frac{ \partial \csa}{\partial \type} d\fals(\ac| \type).
\end{align}
Note that if $Q$ can take any value in $\mathbb{R}$ then it is analogous to a transfer making our problem similar to a mechanism design problem with quasilinear payoffs. \lemmaref{lem:IC} that follows  is standard.
\begin{Lem}\label{lem:IC}A mechanism $\aloc,  \fals$ satisfies truth-telling $\iff$
\begin{enumerate} 
\item $C(\type)$ is increasing 
\item $C(\type)$ belongs to the subgradient of $U$
\item $U(\type)=U(\smin)+\int_{\smin}^{\type}C(z)dz=U(\smax)-\int_{\type}^{\smax}C(z)dz$.
\end{enumerate} 
\end{Lem}
In our setting, however, $Q$ can only take values in $[0,1]$. We proceed to build and solve a relaxed problem and in the process ensure that the $Q$ is in the correct range namely in $[0,1]$. 
Combining \eqref{eq-u-agent} and the equality in item 3 above we can express $Q$ as follows:
\begin{align}\label{eq-int-rep-new}
Q(\type)&=U(\type)+\int_{\ac\in \Ac}\csa d\fals(\ac| \type)=U(\smin)+\int_{\smin}^{\type}C(z)dz+ \int_{\ac\in \Ac}  \csa d\fals(\ac| \type) \notag\\
&=U(\smin) -\int_{\smin}^{\type}\left[ \int_{\ac\in \Ac}  \frac{ \partial \cxa}{\partial z}d\fals(\ac| z)\right]dz+ \int_{\ac\in \Ac}  \csa d\fals(\ac| \type).
\end{align}

Recall that in the case we are analyzing $\mathbb{E}_{F}[\type]<0$. Let $\type_0$ be such that $\int_{\type_0}^{\smax} z dF(z)=0,$ so $\type_0$ is the type above which the conditional expectation of the agent's type is equal to $0$. The designer wants to minimize the approval probability for negative types.
Because $Q(\type)$ is increasing in $U(\type)$ and in falsification costs, at an optimum there is a boundary type $\type_*\geq \type_0$\footnote{We explain why the boundary type must be above $\type_0$ below.} such that (i) $U(\type)=0$ for all $\type \in [\smin,\type_*]$ (ii) $\fals(\ac| \type)=\delta_\type$ for all $\type \in [\smin,\type_*]$; no falsification for these low types sets falsification cost to zero and ensures $Q(\type)=0$. With these observations \eqref{eq-int-rep-new} writes:

\begin{align}\label{eq-int-rep-new-2}
Q(\type)= -\int_{\type_*}^{\type}\left[ \int_{\ac\in \Ac}  \frac{ \partial \cxa}{\partial z}d\fals(\ac| z)\right]dz+ \int_{\ac\in \Ac}  \csa d\fals(\ac| \type).
\end{align}
We employ \eqref{eq-int-rep-new-2} and standard arguments to rewrite the designer's objective of the reduced problem as follows: 
\small
\begin{align*}
\int_{\type_*}^{\smax}Q(\type)\type f(\type)d \type= \int_{\type_*}^{\smax}\int_{\ac\in \Ac}  \left( \cscale c(\ac,\type) \type -\frac{ \partial \cscale c(\ac,\type)}{\partial \type } \mathbb{E}_F[z| z\geq \type ] \frac{\left(1-F(\type ) \right)}{f(\type )}\right)d\fals(\ac| \type )
 f(\type )d\type  
\end{align*}
\normalsize
and the principal's problem becomes: 
\begin{align*}
&\max_{\aloc,\fals } \int_{\type_*}^{\smax}\int_{\ac\in \Ac}  \left( \cscale c(\ac,\type) \type -\frac{ \partial \cscale c(\ac,\type)}{\partial \type } \mathbb{E}_F[z | z\geq \type ] \frac{\left(1-F(\type ) \right)}{f(\type )}\right)d\fals(\ac| \type )
 f(\type )d\type   \\
&\text{ subject to $C$ increasing and $Q\in [0,1]$} 
\end{align*}
where $C$ is defined in \eqref{eq-int-rep-0-new}. 

\paragraph{Relaxed problem} As usual, we solve the relaxed problem ignoring the monotonicity constraint on $C$ which is required for truth-telling. The designer's objective is linear in $ \fals$'s. Moreover, in \lemmaref{LemMonAloc} we have established that the assignment probability $\aloc(\ac,\type)$ is increasing in $\ac$. Then, without loss of generality we can restrict attention to score submission recommendations weakly above the natural score $\type$. The pointwise optimal $\fals$ is by construction deterministic (a Dirac on some \ac) and we denote it simply as $\ac^*: \Types \to \Ac$. This optimal action recommendation solves for each $\type \in \Types$ the following:
\begin{align*}
&\max_{ \ac \in \Ac}
 \cscale c(\ac,\type)\type+\frac{ \partial \cscale c(\ac,\type)}{\partial \type } \mathbb{E}[x | x\geq \type]  \frac{\left(1-F(\type) \right)}{f(\type)}.
\end{align*}

\paragraph{Monotonicity} Per \lemmaref{lem:IC} truth-telling requires that $C(\type)$ is increasing in \type. Using the pointwise optimal recommendation strategy which is deterministic, $C(\type)$ becomes:
\begin{align}\label{eq-int-rep-quad}
C(\type)= -   \frac{ \partial \cscale c(\ac^*(\type),\type)}{\partial \type}.
\end{align}
Note that $C$ is increasing so long as $ -   \frac{ \partial \cscale c(\ac^*(\type),\type)}{\partial \type}$ is  increasing in \type.
%

In what follows, we solve for the pointwise optimal $\fals$ and derive the corresponding optimal $\aloc$ for two classes of falsification cost functions conditions (i) linear in distance falsification costs\footnote{Among others, \cite{li2024screening} assume linear costs.} and (ii) quadraric costs.\footnote{Among others, \cite{frankel2019muddled,frankel2020improving} analyze quadratic costs.} Linear costs are trivially concave. Quadratic costs are convex but the solution approach applies to this case as well as we explain below. In the linear cost case, the solution to the relaxed program is feasible for all distributions of scores. For the case of quadratic costs, the solution is feasible for scores distributions $F$ satisfying the monotone hazard rate property. 

\subsubsection{Linear cost}
Suppose $\csa= \cscale | \ac-\type|$. As anticipated above, we solve for the optimal mechanism for the interesting range of $\gamma$ in which the first-best is not feasible. This is the range of gaming abilities such that \eqref{eq:gamma-ob} is satisfied, which for the linear costs becomes $1-\cscale \smax \geq 0$ or 
$\gamma > \smax $.

For $\ac\geq \type$, the cost is $\cscale(\ac-\type)$ and its derivative is $\frac{ \partial \csa}{\partial \type} =-\cscale$ and 
\[
C(\type)= - \int_{\ac\in \Ac}  \frac{ \partial \csa}{\partial \type} d\fals(\ac| \type)=\cscale.
\]
The principal's program for this cost function becomes:
\begin{align*}
&\max_{\aloc,\fals } \int_{\type_*}^{\smax}\int_{\ac\in \Ac}  \left(  \cscale(\ac-\type) \type+\cscale \mathbb{E}[x | x\geq \type]  \frac{\left(1-F(\type) \right)}{f(\type)} \right) d\fals(\ac| \type) f(\type)d\type \\
&\text{ subject to $C$ increasing and $Q \in [0,1]$} 
\end{align*}
and the pointwise optimum solves:
\begin{align*}
&\max_{ \ac \in [\type,\smax]}
 \cscale(\ac-\type)\type+\cscale \mathbb{E}[x | x\geq \type]  \frac{\left(1-F(\type) \right)}{f(\type)}.
\end{align*}
The expression is linear in $\ac$ and the optimum is a corner solution
\begin{align*}
\ac^*( \type)= \begin{cases} \smax \text{ for }\type\geq 0\\
\type \text{ for } \type<0  \end{cases} \quad \text{ and } \quad 
C( \type)= \begin{cases} \cscale \text{ for }\type\geq 0\\
0 \text{ for } \type<0  \end{cases}
\end{align*}
which is increasing. Thus, the monotonicity constraint is satisfied for all type distributions $F$. 


For \smax, $c(\ac^*(\smax)|\smax)=c(\smax|\smax)=0$, so $U(\smax)=Q(\smax)$. We let $Q(\smax)\equiv p^*(\gamma)$  and we proceed to identify its value below. The assignment probability for $t\geq 0$ is 
\begin{align*}
Q(\type)&= U(\type)+c(\ac^*(\type)|\type)\\
&= U(\smax)-\int_{\type}^{\smax}C(x)dx
+c(\ac^*(\type)|\type)\\
&= U(\smax)-\int_{\type}^{\smax}\cscale dx
+c(\ac^*(\type)|\type)\\
&=p^*(\gamma)-\cscale(\smax-\type)+ \cscale(\smax-\type)\\
&=p^*(\gamma)
\end{align*}
whereas for 
$t<0$ we have $c(\ac^*(\type)|\type)=c(\type|\type)=0$ and we obtain:
\begin{align*}
Q(\type)&= U(\smax)-\int_{\type}^{\smax}\cscale dx
+c(\ac^*(\type)|\type)\\
&=p^*(\gamma)-\cscale(\smax-\type).
\end{align*}
We also need to satisfy the boundary conditions, namely 
$U(\type_*)=0 \iff p^*(\gamma)=\cscale(\smax-\type_*)$. In addition, $p^*(\gamma) \leq 1$. Suppose that $\cscale(\smax-\smin)\leq 1$ then the constraint $p^*(\gamma) \leq 1$ does not bind and the optimal value of $t_*$ maximizes
\[ \int_{t_*}^0[p^*(\gamma)-\cscale(\smax-\type)] f(\type)d\type+\int_{0}^{\smax} p^*(\gamma) f(\type)d\type=\int_{t_*}^0[\cscale(\type-\type_*)] f(\type)d\type+\int_{0}^{\smax} \cscale(\smax-\type_*) f(\type)d\type \]
The first-order condition is  $\int_{t_*}^{\smax}f(\type)d\type=0$ which yields $\type_*=\type_0$. Together with the probability constraint with thus obtain:
\begin{align}\label{eq:tstar}
 \type_*=\min\left\{\type\in[\type_0,\smax]\,:\,\frac{1}{\gamma}(\smax-\type)\leq 1\right\}.
 \end{align}
Therefore, whenever $(\smax-\type_0)\leq \gamma$ the probability constraint does not bind and $\type_*=\type_0$ and
\[p^*(\gamma) =\cscale(\smax-\type_0).\]
Else, that is when $(\smax-\type_0)> \gamma$,  we set $p^*=1$ and $\type^*$ satisfies.
Putting everything together:
\begin{align*}
Q^*( \type)= \begin{cases} p^*(\gamma) \text{ for }\type\geq 0\\
p^*(\gamma)-\cscale(\smax-\type) \text{ for } t_*\leq \type<0  \\
0 \text{ otherwise }\end{cases}
\end{align*}
where $p^*(\gamma)=\min\left\{1,\cscale(\smax-\type_0)\right\}$. We have therefore established the following proposition:
\begin{Prop}Suppose that $\csa= \cscale | \ac-\type|$. Then, the optimal mechanism is 
 \begin{align*}
 Q^*(\type)=\aloc^*(\ac^*( \type),\type)=\begin{cases} p^*(\gamma) \text{ for }\ac=\smax\\
[p^*(\gamma)-\cscale(\smax-\type)]^+ \text{ for }  \type<0  
\end{cases} \text{ } \ac^*( \type)= \begin{cases} \smax \text{ for }\type\geq 0\\
\type \text{ for } \type<0  \end{cases}
 \end{align*}
 where $p^*(\gamma)=\min\left\{1,\cscale(\smax-\type_0)\right\}$. When $\gamma \leq \smax$ the first-best is achieved, when $(\smax-\type_0)\geq \gamma \geq \smax$ all positive types are approved with certainty while negatives are randomly approved. Finally, when $\gamma >(\smax-\type_0)$  all positive types are approved with $p^*(\gamma)<1$ while are randomly approved.
\end{Prop}
It is easy to see that the optimal mechanism can be implemented by a test that randomly assigns inputed types (here types are scores) to an approval or rejection recommendation to a  decision-maker. It thus does not require commitment to the decision nor type reports. 
%


\subsection{Quadratic cost}
Suppose $\csa= \cscale (\ac-\type)^2$. The agent's payoff from truth-telling writes:
 \begin{align*}
 U(\type)
 &=\max_{\typeb \in \Types} \left(Q(\typeb)- \int_{\ac\in \Ac}\cscale (\ac^2+\type^2-2\ac\type) d\fals(\ac| \typeb) \right)\\
 &=\max_{\typeb \in \Types} \left(Q(\typeb)- \int_{\ac\in \Ac}\cscale (\ac^2-2\ac\type) d\fals(\ac| \typeb) \right)- \cscale \type^2.
  \end{align*}
Choosing a report \typeb\ to maximize payoff solves the following equivalent problem:
 \begin{align*}
 \tilde{U}(\type)&=\max_{\typeb \in \Types} \left(Q(\typeb)- \int_{\ac\in \Ac}\cscale (\ac^2-2\ac\type) d\fals(\ac| \typeb) \right)
  \end{align*}
with modified cost function 
\begin{align}\label{eq:modfified-cost}
 \tilde{c}(\ac|\type) \equiv \cscale (\ac^2-2\ac\type) 
   \end{align}
 which is linear and thus concave in $\type.$

For $\ac\geq \type$, the  derivative of $ \tilde{c}$ is $-2\cscale\ac$ and 
\[
 \tilde{C}(\type)=2\cscale\ac .
\]
Then, the principal's objective becomes:
\begin{align*}
\hspace{-0.7in}\int_{\type_*}^{\smax}Q(\type)\type f(\type)d \type &  =  \int_{\type_*}^{\smax}\int_{\ac\in \Ac}  \left(  \cscale (\ac^2-2\ac \type) \type+ 2\cscale\ac \mathbb{E}[z | z\geq \type]  \frac{\left(1-F(\type) \right)}{f(\type)} \right) d\fals(\ac| \type) f(\type)d\type\\
&  = \int_{\type_*}^{\smax}\int_{\ac\in \Ac}  \left(  \cscale (\ac^2-2\ac \type) \type f(\type)+ 2\cscale\ac \int_{\type}^{\smax}z f(z)d z \right)d\fals(\ac| \type) d\type
\end{align*}
and the corresponding relaxed problem is:
\begin{align*}
\max_{\ac  \in [t,\smax]}  \cscale (\ac^2-2\ac \type) \type f(\type)+ 2\cscale\ac \left(\int_{\type}^{\smax}z f(z)d z\right).
\end{align*}

Maximizing pointwise as before we obtain the following optimal falsification target:
\begin{align*}
&\ac^*(\type)=\type-\frac{1}{\type f(\type )}\left(\int_{\type}^{\smax}z f(z)d z \right).
\end{align*}


We now show  that whenever $F$ satisfies the monotone hazard rate property, then  $\ac^*$ is increasing in $\type$. This property will be used below to establish monotonicity of $C$ as well as to identify the type above which $Q$ reaches its maximum value. 
\begin{Lem}[Increasing action recommendation]\label{lem:mon-rec} Assume $F$ has nondecreasing hazard rate. Then, the optimal action recommendation of the relaxed problem $a^*$ is increasing in the natural score for all scores $[\type_0,\smax]$.
\end{Lem}
\paragraph{Verifying monotonicity} Incentive compatibility requires that $C(\type)$ is increasing in \type. Using our pointwise optimal recommendation strategy which is deterministic, $C(\type)$ becomes:
\begin{align}\label{eq-int-rep-quad}
 \tilde{C}(\type)= -   \frac{ \partial \tilde{c}(\ac^*(\type)|\type)}{\partial \type}=2\cscale\ac^*(\type)
\end{align}
which is increasing in \type\ as desired under MHR because as \lemmaref{lem:mon-rec} established  $\ac^*$ is increasing. Thus the solution of the relaxed problem satisfies monotonicity. 
 
\paragraph{Identifying the growth interval} We proceed to identify the smallest type at which the approval probability reaches its highest value. Note that for $\type<0$ but very close to $0$, the optimal action $\ac^*$ explodes to infinity so there is some $\type<0$ denoted by $\type^\dagger$ such that $\ac^*(\type^\dagger)=\smax$ therefore because $\ac^*$ is increasing we have
\begin{align}\label{eq:fals-cap}
 \type^\dagger=(\ac^*)^{-1}(\smax).
 \end{align} 
By definition $\type_*$ satisfies $U(\type_*)=0$. From the discussion of the case of linear costs we also know that $U(\smax)=p^*(\gamma)$.
Also,
$U(\type)=\tilde{U}(\type)- \cscale \type^2$ and $U(\smax)+ \cscale \smax^2=\tilde{U}(\smax)$.

Leveraging these equalities and condition 3 of \lemmaref{lem:IC} to express $\tilde{U}$ we obtain:
\begin{align}\label{eq-boundary}
U(\type_*)&=\tilde{U}(\smax)-\int_{\type_*}^{\smax}C(z)dz - \cscale \type_*^2=U(\smax)-\int_{\type_*}^{\smax}C(z)dz +\cscale (\smax^2-\type_*^2)=0.
\end{align}
Following an analogous procedure as we did for linear costs, we can show that when $\gamma >(\smax-\type_0)^2$ the probability constraint does not bind and  $\type_*=\type_0$ and using this value we can pin down $p^*(\gamma)$ because 
\begin{align}\label{eq-boundary-2}
U(\type_0)=0\iff p^*(\gamma)=\int_{\type_0}^{\smax}2\cscale\ac^*(z)dz +\cscale (\smax^2-\type_0^2).
\end{align}
Instead, when $(\smax-\type_0)^2\geq \gamma$, the probability constraint binds so $p^*(\gamma)=1$ which together with $U(\type_*)=0$ pins down $t_*$:
\begin{align}\label{eq-boundary-3}
U(\type_*)&=1-\int_{\type_*}^{\smax}2\cscale\ac^*(z)dz+\cscale (\smax^2-\type_*^2)=0.
\end{align}
Note that condition 3 of \lemmaref{lem:IC} implies that whenever $U(\type_*)=0$ then $U(\type)\geq 0$ for all $\type>\type_*$. Moreover by the construction of the falsification strategy for types below $\type_*$ (namely no falsification and zero assignment probability) we have that the  participation constraints are satisfied for all \type. 
Therefore the pointwise optimal score submission is:

\begin{align*}
\ac^*(\type)=\begin{cases} \type \text{ for } \type < \type_*\\
\type-\frac{1}{\type f(\type)}\left(\int_{\type}^{\smax}z f(z)d z\right) \text{ for } \type \in [\type_*, \type^\dagger) \\
\smax  \text{ for }  \type^\dagger\leq \type
\end{cases}
\end{align*}
where $\type_*$ satisfies \eqref{eq-boundary-3} or it is equal to $\type_0$ and $\type^\dagger$ satisfies \eqref{eq:fals-cap}.

To calculate the assignment probability we use $\ac^*$ and condition 3 of \lemmaref{lem:IC} but now scale back the cost to $c$
\begin{align*}
Q^*(\type)&= U(\type)+c(\ac^*(\type)|\type)\\
&=U(\smax)+\cscale \smax^2-\int_{\type}^{\smax}C(z)dz - \cscale \type^2+\int_{\ac\in \Ac}\cscale (\ac^2+\type^2-2\ac\type) d\fals^*(\ac| \type)\\
&= p^*(\gamma)+\cscale \smax^2-\int_{\type^\dagger}^{\smax}2\cscale dz-\int_{\type}^{\type^\dagger}2\cscale\ac^*(z)dz+\cscale ((\ac^{*}(\type))^2-2\ac^*(\type)\type).
  \end{align*}

For $\gamma > c(\smax,0) $
\begin{align}
Q^{*}(\type)&=\begin{cases} 0 \text{ for } \type < t_*\\
p^*(\gamma)+\cscale \smax^2-\int_{\type^\dagger}^{\smax}2\cscale dz-\int_{\type}^{ \type^\dagger}2\cscale\ac^*(z)dz+\cscale ((\ac^{*}(\type))^2-2\ac^*(\type)\type)
\text{ for } \type \in [\type_*, \type^\dagger) \\
p^*(\gamma)\text{ for } \type\in [ \type^\dagger,\smax]
\end{cases}
\end{align}
While, as anticipated earlier, whenever $\gamma\leq  c(\smax,0)$ the first-best is achieved:
\begin{align}
Q^{\text{*}}(\type)&=\begin{cases} 0 \text{ for } \type < \smax\\
1\text{ for } \type=\smax.
\end{cases}
\end{align}   
We have therefore established the following proposition:
\begin{Prop}Suppose that $\csa=  \cscale (\ac-\type)^2$ and that $F$ satisfies the monotone hazard rate property. Then the optimal mechanism is 
 \begin{align*}
 \aloc^*(\ac^{*}(\type),\type)&=\begin{cases}p^*(\gamma)\text{ for } \type\in [\type^\dagger,\smax]\\
p^*(\gamma)+\cscale \smax^2-\int_{ \type^\dagger}^{\smax}2\cscale dz-\int_{\type}^{\type^\dagger}2\cscale\ac^*(z)dz+\cscale ((\ac^{*}(\type))^2-2\ac^*(\type)\type)
\text{ for } \type \in [\type_*, \type^\dagger) \\  
 0 \text{ for } \type <\type_*
\end{cases}\\
\ac^*(\type)&=\begin{cases} 
\smax  \text{ for }  \type\in [\type^\dagger,\smax]\\
\type-\frac{1}{\type f(\type)}\left(\int_{\type}^{\smax}z f(z)d z\right) \text{ for } \type \in [\type_*,\type^\dagger)\\
\type \text{ for } \type < \type_*\\
\end{cases}
\end{align*}
where $\type_*(\nu)$ satisfies \eqref{eq-boundary-3} or it is equal to $\type_0$ and $\type^\dagger$ satisfies \eqref{eq:fals-cap}. When $\gamma \leq \smax^2$ the first-best is achieved, when $(\smax-\type_0)^2\geq \gamma \geq \smax^2$ then $p^*(\gamma)=1$ all positive types are approved with certainty while negatives are randomly approved with a probability increasing in \type. Finally, when $\gamma >(\smax-\type_0)^2$  all positive types are approved with $p^*(\gamma)=\cscale (\smax-\type_0)^2 <1$  while negatives are randomly approved with a probability increasing in \type.\end{Prop}
As in case of linear costs, it is easy to see that the optimal mechanism can be implemented by a test that randomly assigns submitted scores to an approval or rejection recommendation to a  decision-maker. It thus does not require commitment nor type reports.

\paragraph{Quadratic cost: uniform distribution}
Suppose the $F$ is the uniform on $[-2,1]$.  Then the optimal mechanism is
\begin{align*}
\ac^*(\type)= \begin{cases}\type \text{ for } \type<-1\\
\frac{3\type}{2}-\frac{1}{2\type}\text{ for }-1\leq \type \leq -\frac13 \\
1 \text{ for }-\frac13 \leq \type \end{cases}
\end{align*}
and for $\gamma>1$:
\begin{align}
Q^{\text{*}}(\type)&=\begin{cases} 0 \text{ for } \type < -1\\
p^*(\gamma)-\cscale-\frac{2}{3\gamma}+\cscale\left(\ln|-\frac13|-\ln|\ac|\right)-\frac{1}{6\gamma}+\frac{6\type^2}{\gamma}+\frac{1}{2\gamma \type^2}-\frac{3}{\gamma}
\text{ for } \type \in [-1, -\frac13) \\
p^*(\gamma)
\text{ for } \type \in [-\frac13, \smax)
\end{cases}
\end{align}
whereas for $\gamma\leq1$ we get the first best. The following figure depicts the optimal interim approval probability and the associated cost for two different values of $\gamma$: 

\begin{figure}[h]
\centering
\scalebox{0.8}
{
\begin{tikzpicture}
\begin{axis}[
    xlabel={$s$},
    ylabel={$Q(\type),C(\type)$},
    xmin=-1, xmax=1,
    ymin=-1, ymax=2.5,
    minor tick num=1,
    axis lines = middle,
    legend pos=north west
]

\addplot[domain=0: 1, samples=400, smooth, unbounded coords=jump, blue] {1-1+ 2*x - x^2+(1-x)^2};
\addplot[domain=0: 1, samples=400, smooth, unbounded coords=jump, red] {(1-x)^2};
\end{axis}
\end{tikzpicture}
\quad
\begin{tikzpicture}
\begin{axis}[
    xlabel={$s$},
    ylabel={$Q(\type),C(\type)$},
    xmin=-1, xmax=1,
    ymin=-1, ymax=2.5,
    minor tick num=1,
    axis lines = middle,
    legend pos=north west
]

\addplot[domain=-1: -0.332, samples=400, smooth, unbounded coords=jump, blue] {(1/4)*(2.43-1-2/3-1/6+(3*x^2)/2+(ln(1/3)-ln(abs(x)))-x^2+((3*x)/2-1/(2*x)-x)^2)};
\addplot[domain=-0.333: 1, samples=400, smooth, unbounded coords=jump, blue] {(1/4)*(2.43-1+ 2*x - x^2+(1-x)^2)};
\addplot[domain=-1: -0.33, samples=400, smooth, unbounded coords=jump, red] {0.25*((3*x)/2-1/(2*x)-x)^2};
\addplot[domain=-0.333: 1, samples=400, smooth, unbounded coords=jump, red] {0.25*(1-x)^2};
\end{axis}
\end{tikzpicture}}
\caption{\footnotesize Left panel $\gamma=1$; \hspace{ 1in} Right panel $\gamma=4$}
\end{figure}
The left panel depicts the interim approval probability for $\gamma=1$ which is equal to the first best. The only distortion is in terms of the agent's utility loss due to falsification cost which is given by the difference $Q(\type)-C(\type)$. The left panel depicts the interim approval probability for $\gamma=4$ which involves distortions because worthy types are assigned an object with probability less than 1 and unworthy types are also assigned objects. The only distortion is in terms of the agent's utility loss due to falsification cost which is given by the difference $Q(\type)-C(\type)$. 

\appendix

\section{Proof of \lemmaref{lem-PC-to-OB}}
Consider an IC DRM mechanism $\nu$. 
Let $\tilde{q}$ satisfy
\begin{align}\label{det-mech-1}
\sum_{\aloc \in \Cd }\nu(\aloc,\ac \mid \type)\aloc(x|\ac)= \tilde{q}(x|\ac)\sum_{\aloc \in \Cd }\nu(\aloc,\ac \mid \type). 
\end{align}
Note that $\tilde{q}$ is a valid score-based decision rule, that is, $\tilde{q}$ maps $\Ac$ to $\Delta(X)$. First, $0\leq \tilde{q}(x|s)\leq 1$ because 
\[\tilde{q}(x|\ac) = \frac{\sum_{\aloc \in \Cd }\nu(\aloc,\ac \mid \type)\aloc(x|\ac)}{ \sum_{\aloc \in \Cd }\nu(\aloc,\ac \mid \type)}\]
and $\aloc(x|\ac) \in [0,1]$. Moreover,
\[\sum_{x\in X}\tilde{q}(x|\ac) =\frac{\sum_{x\in X}\sum_{\aloc \in \Cd }\nu(\aloc,\ac \mid \type)\aloc(x|\ac)}{ \sum_{\aloc \in \Cd }\nu(\aloc,\ac \mid \type)}=\frac{\sum_{\aloc \in \Cd }\nu(\aloc,\ac \mid \type)\underbrace{\sum_{x\in X}\aloc(x|\ac)}_{=1}}{ \sum_{\aloc \in \Cd }\nu(\aloc,\ac \mid \type)}=1.\]
Define a new mechanism $\mu= \alpha \circ \fals$ where 
\begin{align}\label{det-mech-2}
\alpha(\typeb)=\delta_{\tilde{q}} \text{ and } \fals(\ac \mid \typeb) \equiv \sum_{\aloc \in \Cd }\nu(\aloc,\ac \mid \type) \quad\  \forall \typeb \in \Types .
\end{align}  
Note that $\fals$ is the marginal of the mechanism $\nu$ over $\Cd$. Observe that the agent's payoff is the same under $\nu$ and $\mu$ because, conditional on each possible report $\typeb \in \Types$, evidence request $\ac \in \Ac$ and for each $x \in X$ we have:
\begin{align*}
& \sum_{\aloc \in \Cd }\nu(\aloc,\ac \mid \type)\aloc(x|\ac)[v(x,\type)-c(\ac,\type)]\\
 &= \tilde{q}(x|\ac)\sum_{\aloc \in \Cd }\nu(\aloc,\ac \mid \type)[v(x,\type)-c(\ac,\type)] \\
 &= \tilde{q}(x|\ac)\fals(\ac \mid \typeb)[v(x,\type) - c(\ac,\type)]
  \\
 &=\mu (\tilde{q},\ac \mid \typeb)[v(x,\type) - c(\ac,\type)]
\end{align*} 
where the first quality uses \eqref{det-mech-1} and \eqref{det-mech-2} and the second and third equalities use \eqref{det-mech-2} and the definition of $\mu$. Summing up over all $x\in X$ and $\ac \in \Ac$ we obtain:
\begin{align*}
U(\type,\typeb,\nu) &=\sum_{x\in X} \sum_{\ac\in \Ac} \sum_{\aloc \in \Cd }\nu(\aloc,\ac \mid \type)\aloc(x|\ac)[v(x,\type)-c(\ac,\type)] \\
 &=\sum_{x\in X} \sum_{\ac\in \Ac}\mu (\tilde{q},\ac \mid \typeb)[v(x,\type)-c(\ac,\type)] \\
 &= U(\type,\typeb,\mu)
\end{align*}  
where $U(\type,\typeb, \cdot) $ is defined in \eqref{eq:lie-payoff}.

\section{Proof of \propref{prop-score}}
Fix an incentive compatible DRM that has a deterministic score recommendation rule.  Consider two types \type\ and \typeb\ with $\fals(\cdot |\type)= \fals(\cdot |\typeb)=\delta_\ac$. Suppose that $\alocm(x | \ac,\type)\neq \alocm(x | \ac,\typeb)$ for some $x$. Truth-telling implies:
\begin{align*}
\sum_{x}\alocm(x | \ac,\type) \beta(\ac,\type)v(x) &\geq \sum_{x}\alocm(x | \ac,\typeb) \beta(\ac,\type)v(x) \iff  \sum_{x}\alocm(x | \ac,\type)v(x) \geq \sum_{x}\alocm(x | \ac,\typeb) v(x)\end{align*}
where the equivalence follows because $ \beta(\ac,\type)>0$. Analogously we obtain:
\begin{align*}
\sum_{x}\alocm(x | \ac,\typeb) \beta(\ac,\typeb)v(x) &\geq \sum_{x}\alocm(x | \ac,\type) \beta(\ac,\typeb)v(x) \iff \sum_{x}\alocm(x | \ac,\typeb) v(x) \geq \sum_{x}\alocm(x | \ac,\type) v(x)
\end{align*}
Combining results to 
\begin{align*}
\sum_{x}\alocm(x | \ac,\typeb) v(x)=\sum_{x}\alocm(x | \ac,\type) v(x) \iff \sum_{x}[\alocm(x | \ac,\typeb)-\alocm(x | \ac,\type)]v(x)=0
\end{align*}
The last inequality implies that the vectors $\alocm(x | \ac,\type)-\alocm(x | \ac,\typeb)$ and $v(x)$ are orthogonal. We also know $\sum_{x}[\alocm(x | \ac,\typeb)-\alocm(x | \ac,\type)]\bf{1}=0$. Letting $\lambda(x)=\alocm(x | \ac,\type)-\alocm(x | \ac,\typeb)$, we summarize the above two observations as follows: 
\begin{align}\label{eq-orth}
\sum_{x}\lambda(x){\bf1}=0 \quad \sum_{x}\lambda(x)v(x)=0.
\end{align}
Now consider the principal and suppose 
\begin{align*}
&\sum_{x}\alocm(x | \ac,\type)[w(\ac,\type)r(x,\ac)+y(\ac,\type)v(x)+\ell(\ac,\type)] \geq \sum_{x}\alocm(x | \ac,\typeb)[w(\ac,\type)r(x,\ac)+y(\ac,\type)v(x)+\ell(\ac,\type)] \iff \\
&\sum_{x}\alocm(x | \ac,\type)[w(\ac,\type)r(x,\ac)+y(\ac,\type)v(x)] \geq \sum_{x}\alocm(x | \ac,\typeb)[w(\ac,\type)r(x,\ac)+y(\ac,\type)v(x)] \iff \\
&\sum_{x}[\alocm(x | \ac,\type)-\alocm(x | \ac,\typeb)][w(\ac,\type)r(x,\ac)+y(\ac,\type)v(x)] \geq 0 \\
&\sum_{x}[\alocm(x | \ac,\type)-\alocm(x | \ac,\typeb)]\left[r(x,\ac)+\frac{y(\ac,\type)}{w(\ac,\type)}v(x)\right] \geq 0 \\
&\sum_{x}[\alocm(x | \ac,\type)-\alocm(x | \ac,\typeb)][r(x,\ac)+z(\ac,\type)v(x)] \geq 0
\end{align*}
where the first implication uses the first equality in \eqref{eq-orth}, the second is a simple rewriting, the third uses the fact that $w(\ac,\type)>0$ and in the fourth we let $z(\ac,\type) \equiv \frac{y(\ac,\type)}{w(\ac,\type)}$.
The above final inequality writes as
\begin{align*}
\sum_{x}[\lambda(x)r(x,\ac)+z(\ac,\type)\lambda(x)v(x)] &\geq 0 \iff
\sum_{x}\lambda(x)r(x,\ac)\geq 0
\end{align*}
where the equivalence uses \eqref{eq-orth}. The final inequality implies that the principal ranks the lotteries $\alocm(\cdot | \ac,\type)$ and $\alocm(\cdot | \ac,\typeb)$ in the same way no matter the agent's type realization. And because agent types are indifferent we can offer the principal's preferred lottery that only depends on \ac\  and we call it $\aloc(\cdot | \ac)$. 
%

\section{Proof of \lemmaref{LemMonAloc}}
Consider a type $\type$ and all scores on the support of $\type$'s falsification strategy. Denote them $\Ac(\type)$. Suppose, for simplicity,  $\Ac(\type)$ is finite. The idea extends to arbitrary supports. Rank its elements from smallest to largest and label so that 
\[ \ac_1 < \ac_2<\dots<\ac_n\]
relabel the corresponding assignment probabilities so that
\[ \aloc_i = \aloc(\ac_i,\type) \]
and falsification requests so that 
\[ \fals_i = \fals(\ac_i,\type) \]
Trivially, this relabelling by construction ensures  the payoff to $\type$ and to mimicking $\type'$ is the same. To see this note:
\begin{align*}
 Q(\type) = \sum_{i=1}^{n}\fals_i \aloc_i = \sum_{i=1}^{n}  \fals(\ac_i,\type)a(\ac_i,\type)=\sum_{\ac\in \Ac(\type)}  \fals(\ac,\type)a(s,t) \text{ and } \\
C(t) = \sum_{i=1}^{n}\fals_i c(\ac_i,\type) = \sum_{i=1}^{n}  \fals(\ac_i,\type)c(\ac_i,\type)=\sum_{\ac\in \Ac(\type)}  \fals(\ac,\type)c(\ac,\type) \text{ and } \\
C(t\mid t') = \sum_{i=1}^{n}\fals_i c(\ac_i,\type') = \sum_{i=1}^{n}  \fals(\ac_i,\type)c(\ac_i,\type')=\sum_{\ac\in \Ac(\type)}  \fals(\ac,\type)c(\ac,\type') .
\end{align*}
If $\aloc_1\geq 1$, then because $\aloc_1\leq 1$ this is only possible when $\aloc_i=1 \forall i \in \{1,\dots,n\}$ hence the approval probability is increasing in $s$ and there is nothing to prove. If the approval probability is not increasing, then 
\begin{align}\label{eq1-mon-lemma}
\aloc_1<1.
\end{align}
We construct an increasing payoff-equivalent approval probability as follows:
\begin{align}\label{eq2-mon-lemma}
\tilde{\aloc}_i= c(\ac_i,\type) \forall i <n \text{ and } \tilde{\aloc}_n=\frac{\fals_n\aloc_n+\sum_{i \neq n}\aloc_i\fals_i-\sum_{i \neq n}c(\ac_i,\type)_i\fals_i}{\fals_n}
\end{align}
\begin{align}\label{eq3-mon-lemma}
\fals_n\tilde{\aloc}_n+\sum_{i \neq n}c(\ac_i,\type)_i\fals_i=\fals_n\aloc_n+\sum_{i \neq n}\aloc_i\fals_i \geq \sum_{i=1}^{n}  \fals(\ac_i,\type)c(\ac_i,\type)=\sum_{\ac\in \Ac(\type)}  \fals(\ac,\type)c(\ac,\type)
\end{align}
where the last inequality follows from obedience that requires
\[\aloc_i - c(\ac_i,\type)\geq 0 \forall i .\]
Then, for \eqref{eq3-mon-lemma} to hold, it must be the case that $\tilde{\aloc}_n\geq c(\ac_n,\type)$ which ensures obedience and that $\tilde{\aloc}$ is increasing given that $c(\ac,\type)$ is increasing in $s$ which implies that 
\[\tilde{\aloc}_{n} \geq c(\ac_n,\type)\geq c(s_{n-1},t)=\tilde{\aloc}_{n-1}\geq \dots c(s_{1},t)=\tilde{\aloc}_{1}.\]
The problem is that it is possible that $\tilde{\aloc}_{n}>1$. In that case we modify the construction as follows:
Set $\hat{a}_{n}=1$ and assigning the difference to lower scores:
\[ d_n\equiv \frac{\fals_n\aloc_n+\sum_{i \neq n}\aloc_i\fals_i-\sum_{i \neq n}c(\ac_i,\type)_i\fals_i}{\fals_n} -1.\]
We use this equivalent expression:
\[ \fals_n d_n = \fals_n\aloc_n+\sum_{i \neq n}\aloc_i\fals_i-\sum_{i \neq n}c(\ac_i,\type)_i\fals_i- \fals_n\]
to increase $\tilde{\aloc}_{n-1}$ to $\hat{a}_{n-1}=\tilde{\aloc}_{n-1}+d_{n-1}$ where $d_{n-1}$ satisfies 
\[  d_{n-1}\fals_{n-1}=  d_{n}\fals_{n}. \]
If $\hat{a}_{n-1}\leq1$. We are done. Otherwise, 
continue in this way. At some point we will stop because \eqref{eq1-mon-lemma} implies that we cannot have all 1's.  
The resulting assignment is increasing, satisfies OB because all assignment probabilities assigned to a score are by construction higher than the cost type $t$ incurs to generate that score. Also,
\begin{align*}& \fals_n +  d_{n-1}\fals_{n-1}+ \sum_{i \neq n}\fals_i c(s_t |t)\\
&= \fals_n +  d_{n}\fals_{n}+ \sum_{i \neq n}\fals_i c(s_t |t) \\
&= \fals_n +   \fals_n\aloc_n+\sum_{i \neq n}\aloc_i\fals_i-\sum_{i \neq n}c(\ac_i,\type)_i\fals_i- \fals_n+ \sum_{i \neq n}\fals_i c(s_i ,t) \\
&= \fals_n\aloc_n+\sum_{i \neq n}\aloc_i\fals_i-\sum_{i \neq n}c(\ac_i,\type)\fals_i+ \sum_{i \neq n}\fals_i c(s_i ,t)\\
&=Q(t).
\end{align*}
Hence the modification results to the same expected approval probability and the same falsification costs for all types and it is payoff equivalent for the agent and the designer.

\section{Proof of \lemmaref{lem:mon-rec}}
Differentiating $\ac^*$ results to:
\begin{align*}
\frac{\partial \ac^*(\type )}{\partial \type }&=2 +\frac{1}{\type ^2f(\type )}\left(\int_{\type }^{\smax }zf(z)d z\right)+ \frac{f'(\type )}{\type (f(\type ))^2}\left(\int_{\type }^{\smax}z f(z)d z \right) \\
&\geq 2 +\frac{1}{\type ^2f(\type )}\left(\int_{\type }^{\smax}z f(z)d z \right)+ \frac{f'(\type )}{\type (f(\type ))^2}\left(\int_{\type }^{\smax}\type  f(z)dz \right)\\
&= 2 +\frac{1}{\type ^2f(\type )}\left(\int_{\type }^{\smax}z f(z)d z \right)+f'(\type )\frac{1-F(\type )}{(f(\type ))^2}\end{align*}
where the inequality follows because $\type \leq z$. Now for $\type \in [\type_0,\smax]$, $\mathbb{E}[z |z\geq \type ]\geq 0$, which implies that only the last term, namely $f'(\type )\frac{1-F(\type )}{(f(\type ))^2}$ could be negative. 

Recall that the derivative of the usual virtual valuation, namely $J(\type )=\type -\frac{1-F(\type )}{f(\type )}$ is:
\begin{align*}
&\frac{\partial J(\type )}{\partial \type }=2 +f'(\type )\frac{1-F(\type )}{(f(\type ))^2}.
\end{align*}
A sufficient condition for $J$ to be increasing is that the distribution has nondecreasing hazard rate. More generally, whenever the usual virtual valuation is increasing, the optimal recommended action is increasing. This is because the derivative is $\frac{\partial \ac^*(\type )}{\partial \type }=\frac{\partial J(\type )}{\partial \type }+\frac{1}{\type ^2f(\type )}\left(\int_{\type }^{\smax}z f(z)d z \right) $ and the term we are adding is positive, hence increasing $J$ (which is ensured by MHR) suffices for $\ac^*$ to be increasing.

\setstretch{1.0}

\bibliographystyle{ecta}
\bibliography{scoring,vasiliki-1}

\end{document}